# Direct Observation of Strongly Tilted Dirac Points at General Positions in the Reciprocal Space

Yangsong Ye[1,2], Shijie Kang[1], Jiusi Yu[1,2], Aoning Luo[1], Xiexuan Zhang[1], Yiyi Yao[1], Ken Qin[1], Bo Hou[4], Haitao Li[1,a)], Xiaoxiao Wu[1,3,a)]

[1]*Modern Matter Laboratory and Advanced Materials Thrust, The Hong Kong University of Science and Technology (Guangzhou), Nansha 511466, Guangzhou, China*

[2]*Department of Physics, The Hong Kong University of Science and Technology (HKUST), Hong Kong, China*

[3]*Low Altitude Systems and Economy Research Institute, The Hong Kong University of Science and Technology (Guangzhou), Nansha, Guangzhou, 511400, Guangdong, China*

[4]*Wave Functional Metamaterial Research Facility (WFMRF), The Hong Kong University of Science and Technology (Guangzhou), Nansha, Guangzhou 511400, Guangdong, China*

[a)]Corresponding authors. Electronic mail: haitaoli@hkust-gz.edu.cn (H. Li), xiaoxiaowu@hkust-gz.edu.cn (X. Wu).

**Abstract**

Type-II Dirac points (DPs), which occur at the intersection of strongly tilted and touching energy bands, exhibit many intriguing physical phenomena fundamentally different from the non-tilted type-I counterparts. Over the past decade, their discovery has spurred extensive research into electronic systems and other Bloch-wave systems, such as photonic and phononic crystals. However, current studies typically focus on type-II DPs along high-symmetry directions in the first Brillouin zone (FBZ) under mirror symmetry conditions, which are highly restrictive and limit further investigations and applications. To overcome the stringent constraint, here we identify and demonstrate the emergence of type-II DPs at general positions inside the FBZ without requiring the mirror symmetry. The type-II DPs, being accidental degeneracies, are experimentally realized on a metacrystal slab with H-shaped metallic patterns. Our findings indicate that even in the absence of mirror symmetry, type-II DPs can emerge at designated locations inside the FBZ by simply rotating the H-shaped patterns and adjusting geometrical and physical parameters. Furthermore, based on the rotated type-II DPs, off-axis conical diffractions have been realized and experimentally observed. Meanwhile, we discovered that during the rotation process, the type-II DPs transform into off-axis type-I DPs, but still strongly tilted, resulting in the emergence of negative refractions. Hence, the generic method we propose for inducing type-II or strongly tilted type-I DPs without the high-symmetry limitations opens potential avenues for related research. For example, the observed off-axis conical diffraction and negative refraction could inspire future development and applications in photonics and other Bloch-wave systems.

**Introduction**

Type-II Dirac points (DP), distinguished by a strong tilt of the Dirac cone which significantly breaks the Lorentz invariance, were first predicted and observed in electronic materials such as strained graphene[1,2]. They exhibit many intriguing physical phenomena that are fundamentally different from the type-I counterparts. Later, type-II DPs are also discovered in photonic systems, including photonic crystals and metacrystals[3,4]. The research into type-II DPs within topological photonics is driven by the quest for exotic phenomena[5-10], in particular their strongly anisotropic responses. The realization of type-II DPs traditionally relies on the mirror symmetry[11-13], which is crucial for the intersection of bands. Therefore, current studies focus primarily on the emergence of type-II DPs along high-symmetry directions[14-16] in the FBZ. Correspondingly, the band structure is conventionally examined along the high-symmetry directions[16,17], leaving the interior of the FBZ largely unexplored. Notable properties such as enhancement of light-matter interaction[18-20], anti-chiral edge state [21-23], and conical diffraction[24-26] are also studied primarily with type-II DPs located on high-symmetry directions, leaving many unknowns regarding their behaviors inside the FBZ. This gap significantly hinders thorough understanding of strongly tilted DPs and their applications. Nevertheless, it appears nearly impossible to realize a type-II DP without the protection provided by mirror symmetry, which is only present along high-symmetry directions in the FBZ.

In this Letter, we demonstrate a straightforward method to achieve accidental degeneracy[27-29] for type-II DPs inside the FBZ without requiring the mirror symmetry to address the challenge. This approach enables the emergence of type-II DPs at general locations inside the FBZ and is experimentally implemented in a metacrystal composed of H-shaped metallic patterns deposited on a dielectric substrate. Following our scheme, type-II DPs can be positioned inside the FBZ through rotation of the H-shaped patterns, with the rotation angle of the DPs being essentially identical to that of the patterns. In experiments, the iso-frequency contour (IFC) around various DP locations are determined through near-field mapping[30], where the first two photonic bands linearly touch each other. Notably, the ability to designate the location of type-II DPs allows for the realization of off-axis conical diffraction, with observed split diffraction beams. Moreover, we discover that throughout the rotation of the H-shaped patterns, type-II DPs could transform into strongly tilted off-axis type-I DPs, and give rise to negative refractions.

Our generic method for generating type-II or strongly tilted type-I DPs without location constraints paves a potential pathway for researching nodal points within the FBZ, including Dirac points and Weyl points, thereby advancing the fields of topological photonics and material science.

**Results**

**1. Emergence of off-axis type-II DPs**

The geometric details of our design are illustrated in FIG. 1(a), where the substrate is a nonmagnetic dielectric material with a relative permittivity of 16. The metallic patterns are fabricated from a copper layer with a thickness of 0.035 mm, as depicted in the side view, while $\beta$ denotes the local rotation angle of the H-shaped pattern with respect to the center of the unit cell. The full-wave three-dimensional (3D) simulations provide insights into the band structure and electric field distribution (See Note 1 of SI for general simulation settings using COMSOL Multiphysics). FIG. 1(b) presents the band structure along the high-symmetry direction for $\beta = 0°$, where the first two bands intersect at 10.5 GHz, forming a strongly titled degenerate point (type-II DP) on the k-path between Γ to X. Correspondingly, FIG. 1(c) displays the electric field ($E_z$) distributions of the eigenmodes of the type-II DP, highlighting that they are the two fundamental electric dipole modes ($P_x$ and $P_y$). In fact, they give rise to the lowest longitudinal and transverse bands of the metacrystal.

While such type-II DPs along high-symmetry directions has been studied, their emergence within the FBZ has not been observed. Unexpectedly, by rotating the H-shaped metallic pattern with an angle of 10°, the simulated band structure along the k-path at a 10° rotation with respect to ΓX shows an intersection of the two bands, as shown in FIG. 1(d). This strongly tilted intersection leads to the formation of an off-axis type-II DP also at 10.5 GHz. Furthermore, rotating the H-shaped metallic pattern by 20° results in a similar band intersection, as shown in FIG. 1(e), with off-axis type-II DPs emerging at 10.5 GHz. The positions of the type-II DPs in the FBZ indicates that rotating the H-shaped metallic pattern causes the type-II DPs to follow a circular trajectory with a largely fixed radius, ensuring that the angle of rotation in real space corresponds consistently with that in k-space. Moreover, by adjusting the geometrical and physical parameters of the dielectric slab and metallic pattern, the intersection frequency can also be tuned (see SI Note 2). Thus, through these manipulations, type-II DPs can emerge at desired general off-axis location in the FBZ. In fact, these DPs, as accidental degeneracies, are stabilized by the composite symmetry $C_2T$ that combines the two-fold rotation $C_2$ ($x \rightarrow -x$, $y \rightarrow -y$) and time reversal operation $T$ ($t \rightarrow -t$). In other words, for other metallic patterns to induce the off-axis type-II DPs, the $C_2$ symmetry is generally necessary.

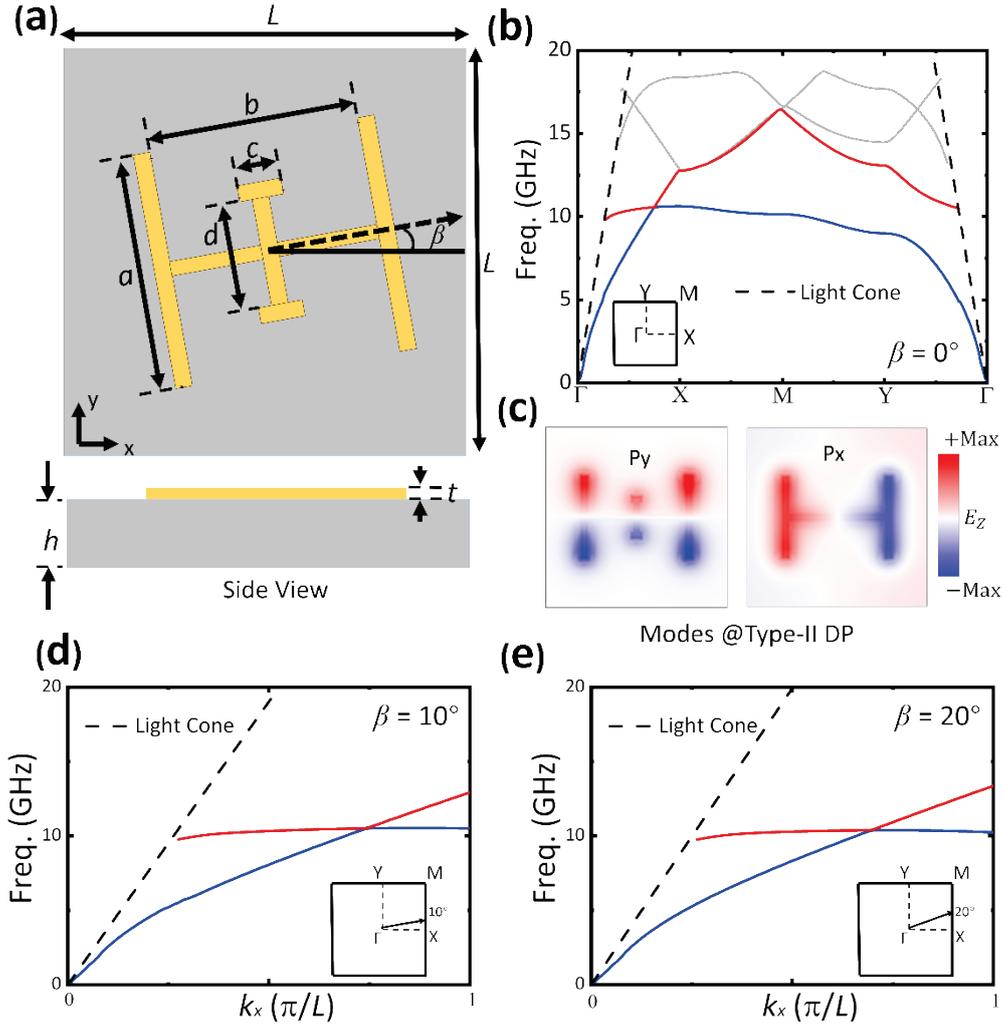

FIG. 1. (a) Illustration of a single unit cell of the metacrystal, featuring a rotated metallic pattern. The dimensions are as follows: $L$ = 4 mm, $a$ = 1.8 mm, $b$ = 2.2 mm, $c$ = 0.2 mm, $d$ = 0.7 mm, $h$ = 2 mm, $t$ = 0.035 mm, with $β$ representing the rotation angle. (b) The simulated band structure along the high symmetry direction for $β$ = 0° shows the blue line (1st band) intersecting the red line (2nd band) at DPs at a frequency of 10.5 GHz along the Γ to X path. The dashed lines indicate the light cone, the grey lines correspond to higher bands, and the inset displays the FBZ. (c)

Electric field distributions ($E_z$) of the two degenerate modes at the Dirac frequency. They are both electric dipole modes, corresponding to the transverse ($P_y$) and longitudinal ($P_x$) bands, respectively. (d) The simulated band structure for a rotation angle of $\beta = 10°$. The inset highlights the calculation path in FBZ. Here, the two bands intersect to form type-II DPs at 10.5 GHz. (e) Similarly, the simulated band structure for $\beta = 20°$, with calculation path as shown in the inset, also featuring type-II DPs at 10.5 GHz.

The **k·p** theory is used to illustrate the dispersion around the DPs[31,32] in the k-space. The effective Hamiltonian around the type-II DPs can be approximated as follows[33]:

$$H = \omega_x \delta k_x \sigma_0 + \omega_y \delta k_y \sigma_0 + v_x \delta k_x \sigma_x + v_y \delta k_y \sigma_y \qquad (1)$$

in which $\boldsymbol{\omega} = (\omega_x, \omega_y)$ refer to the tilt of Dirac cone along the $x$ and $y$ directions, and $\delta \mathbf{k} = (\delta k_x, \delta k_y)$ is the Bloch wave vector deviated from the DP. The $\sigma_0$ represents the 2×2 identity matrix, $\sigma_i$ ($i = x, y, z$) are the Pauli matrixes, and the parameters $v_x$ and $v_y$ represent the cone angles along the $x$ and $y$ directions, respectively. From the cone angle $\mathbf{v} = (v_x, v_y)$ and the tilt of Dirac cone $\boldsymbol{\omega}$, the tilt parameter $\mu$ is defined as:

$$\mu = \sqrt{\left(\frac{\omega_x}{v_x}\right)^2 + \left(\frac{\omega_y}{v_y}\right)^2} \qquad (2)$$

According to the value of $\mu$, DPs can be divided into type-I ($\mu < 1$) and type-II ($\mu > 1$). In addition, from the Eq. [1], the dispersion relation around the DPs can be analytically expressed as:

$$f_{\pm} = \frac{1}{2\pi}(\omega_x \delta k_x + \omega_y \delta k_y \pm \sqrt{v_x^2 \delta k_x^2 + v_y^2 \delta k_y^2}) \qquad (3)$$

where the sign of ± refers the upper (+) and lower (−) band. The degenerate DP is marked as $\delta k_x = \delta k_y = 0$. The analytic dispersions around the type-II DPs are fitted and their iso-frequency contours at the Dirac frequency (10.5 GHz) are evaluated and plotted, as shown in Note 3 of SI, which indicate the emergence of conical diffractions.

## 2. Band structures and emergence of off-axis conical diffraction

The tilted band structure dictates that conical diffraction arises in conjunction with type-II DPs. Conical diffraction is characterized by its unique properties: a flat wavefront akin to that of the incident wave and a conical pattern of the energy flows represented by Poynting vectors. To excite type-II DPs at the specific location, thereby facilitating the on- and off-axis conical diffraction, dielectric waveguides are designed to operate at first fundamental mode with their dispersions matching those of the type-II DPs. Consequently, the rotation angle of the rectangular waveguide must be consistent with the rotation angle of the metallic pattern, as illustrated in FIG. 2(a).

Type-II DPs are typically protected by a combination of spatial symmetries, particularly mirror symmetries. In our instance, a mirror symmetry is not essential for the formation of type-II DPs. Even though rotational operations disrupt the mirror symmetry, in our metacrystal, type-II DPs persist and rotate in associated with the metallic pattern. In fact, whether the metallic pattern is rotated or not, an accidental degeneracy emerges between the 1st and 2nd bands along the rotated path in the k-space, as demonstrated in FIG. 2(b), (c), and (d). The strongly titled dispersions, both positive, give rise to a type-II DP, with the other one guaranteed by the time-reversal symmetry. It is noteworthy that the radius between the location of the type-II DPs and the origin in k-space remains largely constant during rotation. Moreover, the eigenfrequency at the type-II DPs is consistently around 10.5 GHz under the geometric configuration specified in FIG. 1(a)..

To verify the on- and off-axis conical diffraction, three primary types of metacrystals are constructed for $\beta = 0°$, $10°$, and $20°$, each with the corresponding excitation dielectric waveguide as shown in FIG. 2(a). FIG.2 (b), (c), and (d) depict the 3D band structure for $\beta = 0°$, $10°$, and $20°$ and the IFCs around the DPs, respectively. In fact, whether the metallic patter is rotated or not, an accidental degeneracy emerges between the 1st and 2nd bands along the rotated path in the k-space. The strongly titled and linearly touching dispersions give rise to a type-II DP, with the other one guaranteed by the time-reversal symmetry. This indicates that the type-II DPs, being accidental degeneracies, persist under rotation regardless of the $\beta$ value. After exciting the DPs in k-space, whether on-axis ($\beta = 0°$) or off-axis ($\beta = 10°$ and $20°$), a conical diffraction is observed at different angles, as shown in FIG. 2(e).

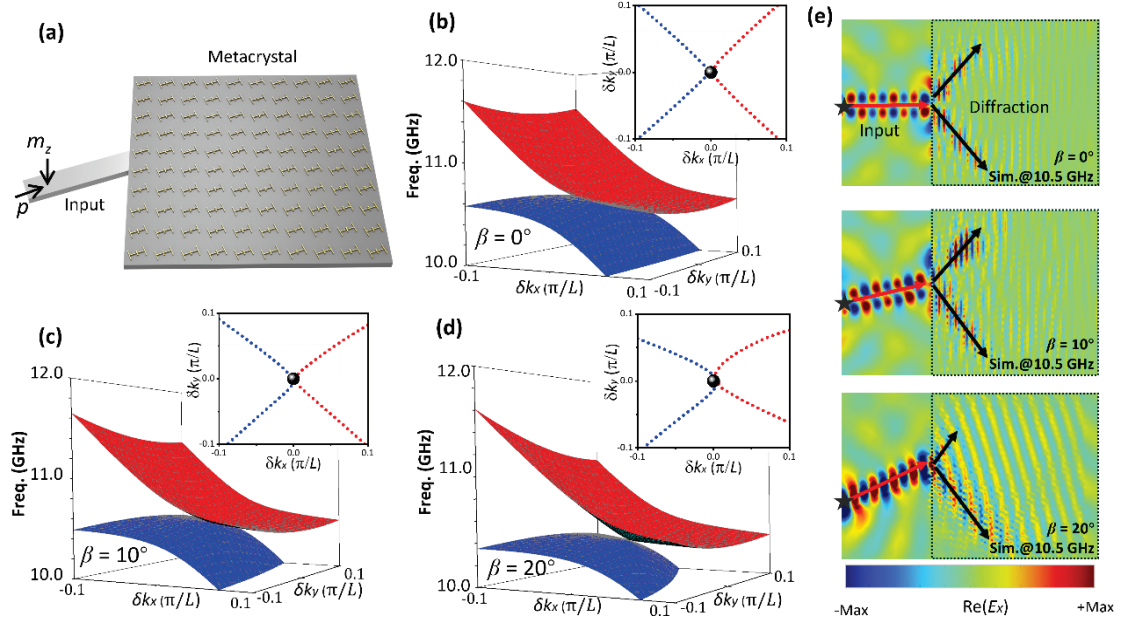

FIG. 2. (a) A schematic illustration of the metacrystal, alongside a tilted input waveguide. An in-plane electric dipole and an out-of-plane magnetic dipole are used to excite the dielectric waveguide. (b)-(d) The numerically calculated 3D band structures for $\beta = 0°$, $10°$, and $20°$, respectively, shown in terms of $\delta k_x$ and $\delta k_x$, measured from the type-II DPs. The blue (red) curved surface represents the 1st (2nd) band. Their intersection defines the type-II DPs. The inset displays the iso-frequency contour at the Dirac frequency. (e) Simulated electric field distribution showing the conical diffractions at $\beta = 0°$, $10°$, and $20°$, respectively. The dielectric waveguide is tilted accordingly. The stars indicate the excitation point at the input waveguide.

## 3. Experimental observation of off-axis type-II DPs.

To demonstrate the type-II DPs at off-axis locations, an electric dipole source together with a magnetic dipole source were employed to excite the first two modes within the frequency range of 8 to 12 GHz. Upon rotating the H-shaped metallic pattern, the type-II DPs in k-space are rotated by an angle equal to that of the H-shaped metallic pattern. To validate the simulation results, metacrystal samples were fabricated according to the geometrical specifications outlined in FIG. 1(a), with dimensions of 180 × 180 mm$^2$ comprising 45 × 45 unit cells. Copper metallic patterns, 18-μm thick, were printed on the dielectric slab and covered with a thin layer of tin to prevent oxidation. The substrates are TP-2 dielectrics, characterized by a relative permittivity of 16 and a tangential loss of 0.001 around 10 GHz. The near-field scanning method was utilized to measure the electric field distribution on a near-field platform (See Note 4 of SI), enabling the identification of type-II DPs in k-space through Fourier transforms. The probing antenna sequentially scanned above the metacrystals at an approximate distance of 1 mm, covering an area of 170 × 170 mm$^2$ with a scanning resolution of 2 × 2 mm$^2$. Fourier transforms were performed on the measured real-space $S_{21}$ data, yielding a total resolution of 170 × 170 points in the experimental k-space data, which spans four Brillouin zones (BZs). Folding the k-space data back to the FBZ, and following the corresponding high-symmetry paths, the measured Fourier spectra for $\beta$ = 0°, 10°, 20°, and 40° are obtained, as depicted in FIG. 3. A cross-shaped pattern of two bright strips, indicative of a tilted intersection between the first two bands, is observed for all these rotations of 0°, 10°, 20°, and 40° in the Fourier spectra. Therefore, the experimental observation of type-II DPs at various general locations in the k-space has been successfully achieved through the fabricated metacrystals.

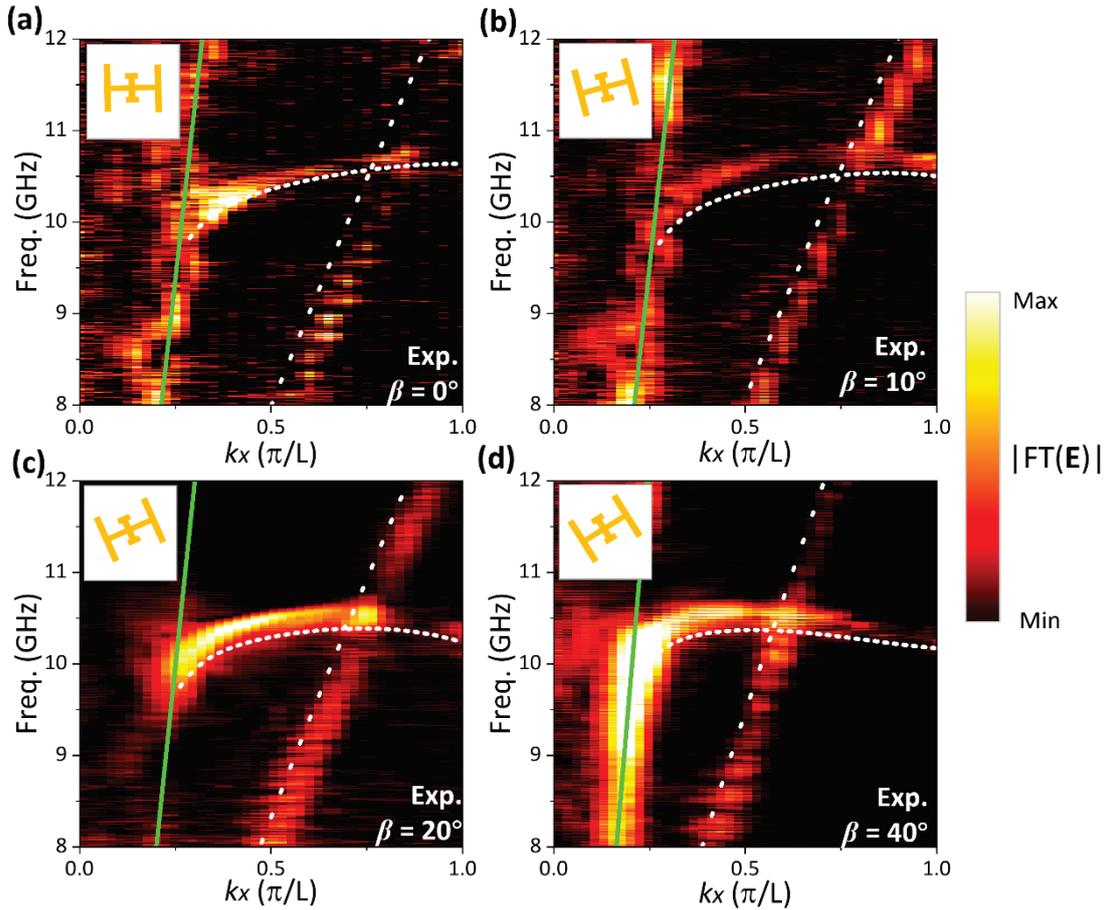

FIG. 3. The experimental Fourier spectra along specific directions within the FBZ, where white dots indicate numerical dispersion and the green solid line represents the light cone, with DPs emerging at 10.5 GHz. (a) Along the Γ-X axis with rotation angle $\beta$ = 0°; (b) Along the 10° path, as indicated in FIG. 1(d), with $\beta = 10°$; (c) Along the 20° path, as depicted in FIG. 1(e), with $\beta = 20°$; and (d) Along the 40° path, a direction similar to those shown in FIG. 1(d) and (e), with $\beta = 40°$. Insets illustrate the corresponding rotated metallic pattern for each spectrum. The Fourier amplitudes in higher BZs are enhanced (×10) for better visualization.

## 4. Experimental observation of conical diffraction.

Three sets of metacrystal samples and dielectric waveguides were fabricated with $\beta$ values of 0°, 10°, and 20° to investigate both on-axis and off-axis conical diffraction phenomena. The samples consist of 45 × 45-unit cells, each measuring 180 × 180 mm$^2$, and the waveguide is tilted according to the varying $\beta$ values relative to these samples. In detail, the rectangular dielectric waveguide has a cross-sectional dimensional of 30 mm × 3 mm, with length being 55 mm. The dielectric waveguide is designed such that its dispersion matches the wave vector and frequency of the strongly DPs. As the $\beta$ value increases, the DPs rotate on the 10.5 GHz plane in k-space, maintaining a largely constant radius value of $0.76\pi/L$. To induce off-axis conical diffraction by exciting DPs at different positions in k-space, the waveguides rotation angle must match $\beta$, as depicted in FIG. 2 (a). The fundamental mode of the waveguide is excited by an electric dipole source together with a magnetic dipole source positioned at its input end. Near-field scanning is performed over a region measuring 170 × 170 mm$^2$ with a resolution of 2 × 2 mm$^2$. FIG. 4(a) and (d) illustrate the experimental field mapping of the real part of the *Ex*-component for $\beta = 10°$ and $\beta = 20°$, respectively, at the Dirac frequency of 10.5 GHz. Similar to the on-axis conical diffraction observed at $\beta = 0°$, the incident wave splits upon entering the metacrystals. The splitting angles for $\beta = 10°$ and $\beta = 20°$ are 102° and 95°, respectively (for the derivation, please refer to Note 5 of SI). The off-axis diffraction pattern rotates anticlockwise by 10° (FIG. 4(a)) and 20° (FIG. 4(d)), corresponding to the rotation of DPs in k-space, in agreement with the simulation results in FIG. 2(e). A Fourier transform is conducted to obtain the Fourier spectra in the reciprocal space, as shown in FIG. 4(b) and (e), with $G_x$ representing the unit reciprocal lattice vector along *x* direction. Around the locations corresponding to the type-II DPs but in the second BZs, that is, differed by $G_x$, the Fourier spectra reveal the characterstic "X" shape, thereby verifying the excitation of the type-II DPs. Unlike at $\beta = 0°$, where the DPs are located along the high-symmetry ΓX path, the DPs are rotated by 10° and 20° with respect to the ΓX path, as depicted in FIG. 4(b) and (e). To confirm that the experimental results align well with the simulation, FIG. 4 (c) and (f) depict the comparisons between the simulated and experimental IFCs. The enlarged areas, with a side length of 0.3 π/L, shows that the "X" shaped IFCs retrieved from the simulations closely match the experimental data for both $\beta = 10°$ and $\beta = 20°$. Thus, the off-axis conical diffractions have been experimentally observed by exciting DPs at varied positions in k-space, and they exhibit rotated diffraction beams. For the analysis of the IFC near the Dirac points and the corresponding conical diffraction, see Note 10 of SI.

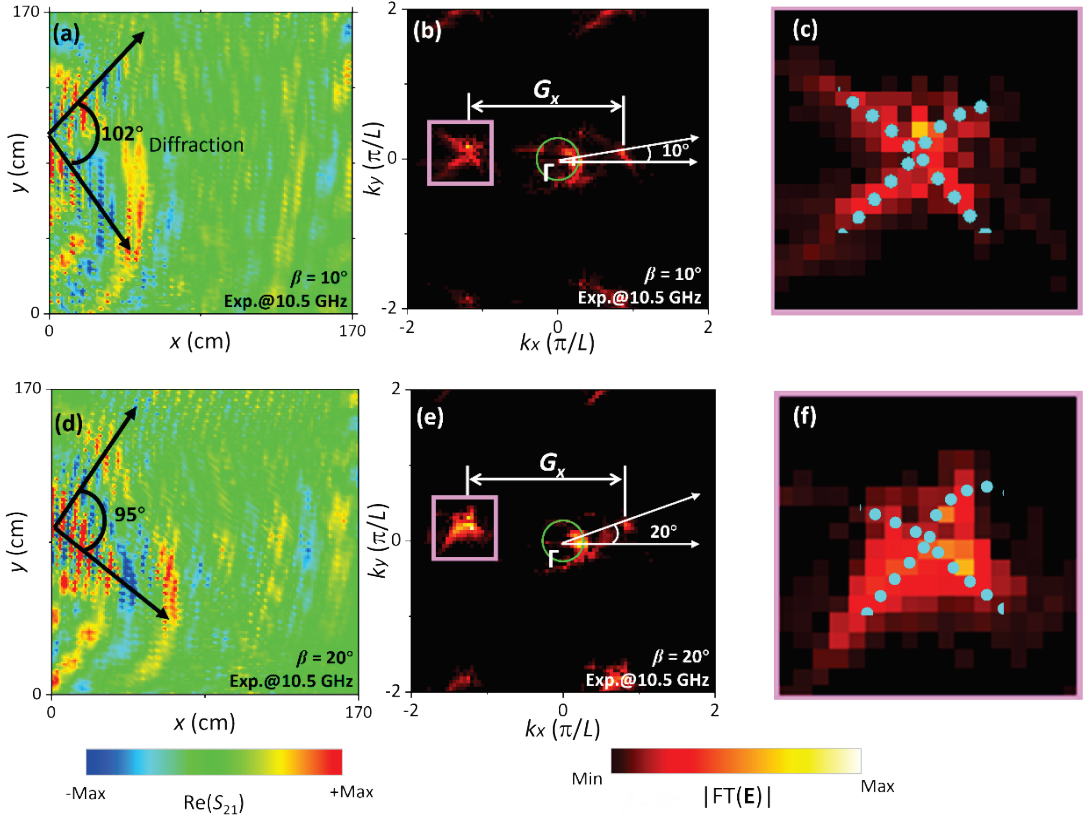

FIG. 4. Experimental observation of off-axis conical diffraction. (a) The measured electric field distribution showing the real part of $E_x$ at 10.5 GHz for $\beta = 10°$ and the splitting angle of 102°. (b) The Fourier spectra for two BZ at the selected frequency of 10.5 GHz for $\beta = 10°$, where bright strips indicate the excited modes and illustrate the IFCs. The green solid circle represents the light cone, the purple solid square indicates the enlarged area, and $G_x$ denotes the unit reciprocal lattice. (c) Enlarged view of the section in (b). The cyan dotted lines represents the numerically calculated IFC with the range $\delta k_x = \delta k_y = \pm 0.15\,\pi/L$. (d) The measured electric field distribution showing the real part of $E_x$ at 10.5 GHz for $\beta = 20°$ and the splitting angle of 95°. (e) Similar to (b), but with $\beta = 20°$. (f) Similar to (c), but corresponding to (e) with $\beta = 20°$.

## 5. Emergence and observation of negative refraction caused by strongly tilted type-I DPs

In our investigation, we have observed that when the parameter $\beta$ ranging from 28° to 62° (for comprehensive details, please refer to the Note 6 of SI), the DPs undergo a transformation from type-II into strongly tilted type-I DP[34]. In this configuration, one of the bands will become notably flat, leading to a transition from conical diffraction to negative refraction[35], as illustrated in FIG. 5(a). To simulate this phenomenon, we use the input frequency at 10.3 GHz and rotate the H-shaped patterns by 40° (for the details of frequency shifting, please refer to Note 7 of SI). The transition to strongly tilted type-I DP is attributed to the near flattening of the longitudinal band. Moreover, the IFC converges into a singular line aligned with the $\beta$ angle or transforms into two curved lines resembling a bifurcation of the single line at the endpoints. This behavior is the consequence of a small mass term in the effective Hamiltonian (See

Note 8 of SI) arising from misalignment of the rotated $P_x$ and $P_y$ electric dipole modes with the square lattice. FIG. 5(b) presents the 3D band diagram and IFCs, revealing the emergence of DPs along the 40° path relative to the ΓX path in the FBZ, with the IFCs at 10.5 GHz around the DPs appearing parallel to the 40° line (See the Supplementary Information for further illustration).

To validate our findings, FIG. 5(c) displays the experimental verification of negative refraction. The experimental results align with the simulation outcomes, demonstrating that the wavefront follows a 40° rotational direction while the energy flow adheres to the negative refraction direction. Moreover, the Fourier spectra from the experimental results are depicted in FIG. 5(d), also showcasing the $k$-space data that includes four BZs around the locations of the strongly tilted type-I DP in the second BZ, the Fourier spectra now reveal a single bright strip instead of an "X" shape. Consequently, the group velocity, determined by $v_g = \partial w/\partial k$, corroborates the negative refraction direction, suggesting consistency between theoretical predictions and experimental observations. Thus, when type-II DPs transition to strongly tilted type-I DP DPs ($\beta$ ranging from 28° to 62°), the off-axis conical diffraction will transform into negative refraction, as we have experimentally observed.

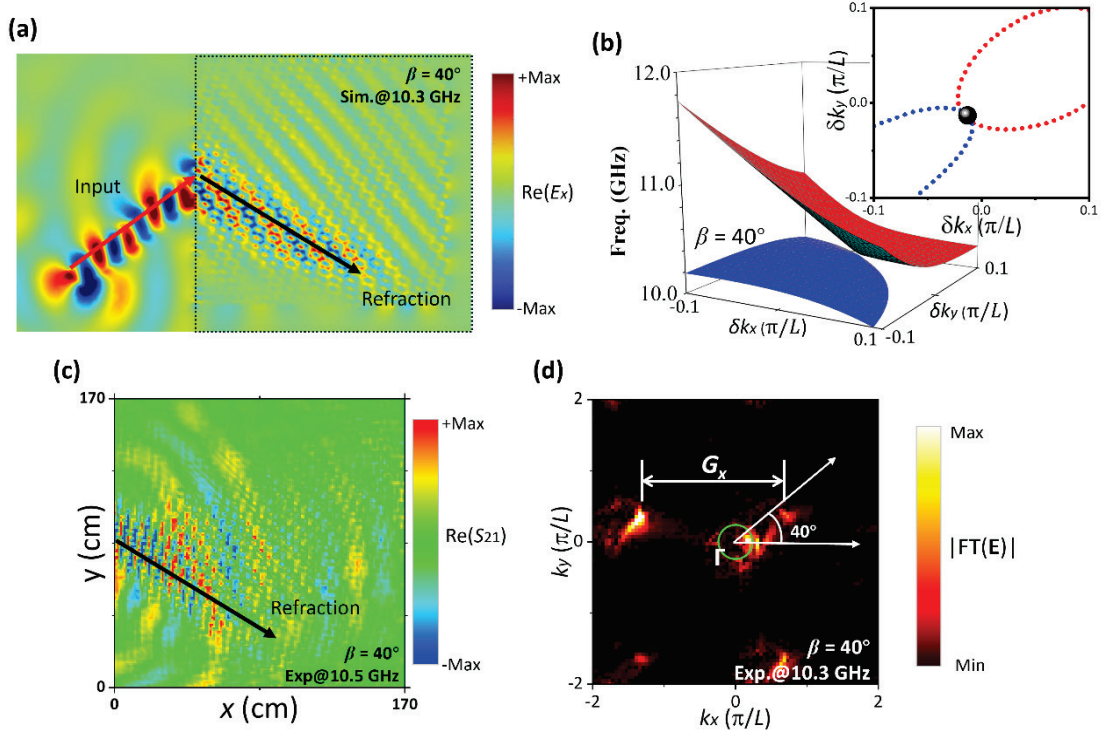

FIG. 5. Strongly-tilted type-I DP and associated negative refraction. (a) Simulated electric field distribution showing negative refraction at $\beta = 40°$, with the dielectric waveguide tilted at 40°. The black dashed square delineates the region of the metacrystal, while the red arrow indicates the direction of the incident wave, and the black arrow represents the direction of refraction. (b) The 3D band diagrams for $\beta = 40°$ are presented in the $k_x$ and $k_y$ plane. The blue curved surface corresponds to the first band, while the red curved surface represents the second band. The intersection of these two surfaces now forms the strongly tilted type-I DP. Additionally, the inset illustrates the IFCs at the Dirac frequency. (c) The measured electric field distribution above the metacrystal sample at the frequency of 10.5 GHz for $\beta = 40°$, corresponding to the square region in (a), showing the negative refraction. (d) The measured Fourier

spectra at the frequency of 10.5 GHz for $β = 40°$ are shown, where bright strips indicate the IFCs. The green solid circle represents the light cone, while $G_x$ denotes the unit reciprocal lattice vector.

**Conclusion**

In conclusion, our research experimentally presents a concise approach for generating strongly tilted DPs through the accidental degeneracy stabilized by the composite *IT* symmetry. Utilizing a metacrystals slab with H-shaped metallic patterns, this method allows for the emergence of type-II DPs or strongly tilted type-I DPs at general location within the k-space, eliminating the need for the stringent mirror symmetry conditions. By simply rotating the H-shaped pattern and adjusting its geometric and physical parameters, we can generally position these strongly tilted DPs inside the FBZ, overcoming the limitations of high-symmetry directions. Moreover, the experimental observations of off-axis conical diffractions and negative refraction further underscore the versatility of our approach. Paving the way for more flexible explorations of type-II and strongly tilted type-I DPs, this advancement not only opens a potential avenue for research of associated phenomena, but also holds notable potential for future applications in photonics and other Bloch-wave systems.

**Acknowledgement**

This work is supported by National Natural Science Foundation of China (No. 12304348, No. 12074279), Guangdong Basic and Applied Basic Research Foundation (No. 2025A1515011470), Guangdong Provincial Project (2023QN10X059), Guangdong University Featured Innovation Program Project (2024KTSCX036), Guangzhou Municipal Science and Technology Project (No. 2024A04J4351), Guangzhou Higher Education Teaching Quality and Teaching Reform Engineering Project (2024YBJG087), HKUST - HKUST(GZ) 20 for 20 Cross-campus Collaborative Research Scheme (G011). X. Wu also acknowledges support from the Modern Matter Laboratory of HKUST(GZ).

**Author Declarations**

The authors have no conflicts to disclose.

**Data Availability**

The data that support the findings of this work are available from the corresponding authors upon reasonable request.

**References**

[1] MO Goerbig, J-N Fuchs, G Montambaux, and F Piéchon, Physical Review B—Condensed Matter and Materials Physics **78** (4), 045415 (2008).

[2] Maximilian Trescher, Björn Sbierski, Piet W Brouwer, and Emil J Bergholtz, Physical Review B **91** (11), 115135 (2015).

3   Wenlong Gao, Biao Yang, Ben Tremain, Hongchao Liu, Qinghua Guo, Lingbo Xia, Alastair P Hibbins, and Shuang Zhang, Nature communications **9** (1), 950 (2018).

4   Sheng Long, Jie Yang, Hanyu Wang, Zhide Yu, Biao Yang, Qinghua Guo, Yuanjiang Xiang, Lingbo Xia, and Shuang Zhang, Optics Letters **48** (9), 2349 (2023).

5   NP Armitage, EJ Mele, and Ashvin Vishwanath, Reviews of Modern Physics **90** (1), 015001 (2018).

6   Arun Bansil, Hsin Lin, and Tanmoy Das, Reviews of Modern Physics **88** (2), 021004 (2016).

7   Ching-Kai Chiu, Jeffrey CY Teo, Andreas P Schnyder, and Shinsei Ryu, Reviews of Modern Physics **88** (3), 035005 (2016).

8   Alex J Frenzel, Christopher C Homes, Quinn D Gibson, YM Shao, Kirk W Post, Aliaksei Charnukha, Robert Joseph Cava, and Dimitri N Basov, Physical Review B **95** (24), 245140 (2017).

9   TE O'Brien, M Diez, and CWJ Beenakker, Physical review letters **116** (23), 236401 (2016).

10  Z. Y. Li, Q. Li, and Z. Li, Chinese Physics B **31** (12) (2022).

11  Xiaoxiao Wu, Xin Li, Ruo-Yang Zhang, Xiao Xiang, Jingxuan Tian, Yingzhou Huang, Shuxia Wang, Bo Hou, Che Ting Chan, and Weijia Wen, Physical Review Letters **124** (7), 075501 (2020).

12  Chuandeng Hu, Zhenyu Li, Rui Tong, Xiaoxiao Wu, Zengzilu Xia, Li Wang, Shanshan Li, Yingzhou Huang, Shuxia Wang, and Bo Hou, Physical Review Letters **121** (2), 024301 (2018).

13  Hai-Xiao Wang, Yige Chen, Zhi Hong Hang, Hae-Young Kee, and Jian-Hua Jiang, npj Quantum Materials **2** (1), 54 (2017).

14  Satoru Hayami, Physical Review B **108** (9), 094416 (2023).

15  Zhi-Yuan Li and Lan-Lan Lin, Physical Review E **67** (4), 046607 (2003).

16  Louis P Bouckaert, Roman Smoluchowski, and Eo Wigner, Physical Review

**50** (1), 58 (1936).

17  D Jetal Chadi and Marvin L Cohen, Physical Review B **8** (12), 5747 (1973).

18  Nicholas Rivera and Ido Kaminer, Nature Reviews Physics **2** (10), 538 (2020).

19  Giuseppe D'Aguanno, Nadia Mattiucci, Claudio Conti, and Mark J Bloemer, Physical Review B—Condensed Matter and Materials Physics **87** (8), 085135 (2013).

20  Erik Petrovish Navarro-Barón, Herbert Vinck-Posada, and Alejandro González-Tudela, ACS Photonics **8** (11), 3209 (2021).

21  Peiheng Zhou, Gui-Geng Liu, Yihao Yang, Yuan-Hang Hu, Sulin Ma, Haoran Xue, Qiang Wang, Longjiang Deng, and Baile Zhang, Physical Review Letters **125** (26), 263603 (2020).

22  Jianfeng Chen, Wenyao Liang, and Zhi-Yuan Li, Physical Review B **101** (21), 214102 (2020).

23  Hongfei Wang, Biye Xie, and Wei Ren, Laser & Photonics Reviews **18** (1), 2300764 (2024).

24  Gunther Schmidt and Bernd H Kleemann, Journal of Modern Optics **58** (5-6), 407 (2011).

25  Jeong-Gil Kim, Chih-Hung Hsieh, Hyungryul J Choi, Jules Gardener, Bipin Singh, Arno Knapitsch, Paul Lecoq, and George Barbastathis, Optics Express **23** (17), 22730 (2015).

26  KE Ballantine, JF Donegan, and PR Eastham, Physical Review A **90** (1), 013803 (2014).

27  Xueqin Huang, Yun Lai, Zhi Hong Hang, Huihuo Zheng, and Che Ting Chan, Nature materials **10** (8), 582 (2011).

28  Lin Xu, Hai-Xiao Wang, Ya-Dong Xu, Huan-Yang Chen, and Jian-Hua Jiang, Optics express **24** (16), 18059 (2016).

29  Ze-Guo Chen, Xu Ni, Ying Wu, Cheng He, Xiao-Chen Sun, Li-Yang Zheng, Ming-Hui Lu, and Yan-Feng Chen, Scientific reports **4** (1), 4613 (2014).

30  N Suresh Kumar, K Chandra Babu Naidu, Prasun Banerjee, TVSRP Anil Babu, and B Venkata Shiva Reddy, Crystals **11** (5), 518 (2021).

31  Xiaoxiao Wu, Yan Meng, Jingxuan Tian, Yingzhou Huang, Hong Xiang, Dezhuan Han, and Weijia Wen, Nature communications **8** (1), 1304 (2017).

32  Jun Mei, Ying Wu, Che Ting Chan, and Zhao-Qing Zhang, Physical Review B—Condensed Matter and Materials Physics **86** (3), 035141 (2012).

33  Si Li, Zhi-Ming Yu, Ying Liu, Shan Guan, Shan-Shan Wang, Xiaoming Zhang, Yugui Yao, and Shengyuan A Yang, Physical Review B **96** (8), 081106 (2017).

34  Dingping Li, Baruch Rosenstein, B Ya Shapiro, and I Shapiro, Physical Review B **95** (9), 094513 (2017).

35  Zhiwei Guo, Haitao Jiang, and Hong Chen, Advanced Photonics Research **2** (1), 2000071 (2021).